\documentclass[a4paper,11pt]{amsart}
\begin{document}

\hyphenation{gra-vi-ta-tio-nal re-la-ti-vi-ty Gaus-sian
re-fe-ren-ce re-la-ti-ve gra-vi-ta-tion Schwarz-schild
ac-cor-dingly gra-vi-ta-tio-nal-ly re-la-ti-vi-stic pro-du-cing
de-ri-va-ti-ve ge-ne-ral ex-pli-citly des-cri-bed ma-the-ma-ti-cal
de-si-gnan-do-si coe-ren-za pro-blem gra-vi-ta-ting geo-de-sic
per-ga-mon cos-mo-lo-gi-cal gra-vity cor-res-pon-ding
de-fi-ni-tion phy-si-ka-li-schen ma-the-ma-ti-sches ge-ra-de
Sze-keres con-si-de-red tra-vel-ling ma-ni-fold re-fe-ren-ces
geo-me-tri-cal}

\title[Schwarzschild manifold \emph{etc.}]
{{\bf Schwarzschild manifold and\\non-regular coordinate
transformations\\(\emph{A critico-historical Note})}}

\author[Angelo Loinger]{Angelo Loinger}
\address{A.L. -- Dipartimento di Fisica, Universit\`a di Milano, Via
Celoria, 16 - 20133 Milano (Italy)}
\author[Tiziana Marsico]{Tiziana Marsico}
%\date{}
\address{T.M. -- Liceo Classico ``G. Berchet'', Via della Commenda, 26 - 20122 Milano (Italy)}
\email{angelo.loinger@mi.infn.it} \email{martiz64@libero.it}
%\thanks{}

\vskip0.50cm

\begin{abstract}
A careful analysis of the maximally extended metrics of
Schwarzschild manifold shows that the \emph{original}
Schwarzschild's solution (1916) and Brillouin's solution (1923)
are the only ones that are adequate from the physical standpoint.
Contrary to the other maximally extended metrics, they represent
faithfully the gravity field created by the mass-point.
\end{abstract}

\maketitle

%%\begin{equation} \label{eq:sevenprime}
%%    \ddot{\Re} + \f\textbf{5}. External
%% {\kappa}{6}\Re \rho=0 , \tag{7'}
%% \end{equation}
%% ``mechanisms'' \textrm{d} \`a
%% \cite{1}
%% eqs.(\ref{eq:six})
%% Schwarzschild


\normalsize

\vskip1.20cm \noindent \textbf{1.} -- According to the current
literature, the maximally extended metrics of Schwarzschild
manifold, which is created by a gravitating point-mass $m$, are
\emph{in primis} the following: \emph{i}) the
Eddington-Finkelstein \cite{1} metric, \emph{ii}) the Lema\"itre
\cite{2} and Robertson \cite{3} metrics, \emph{iii}) the
Kruskal-Szekeres \cite{4} metric.

\par We anticipate their main defects: $\alpha$) the ``soft''
singularity at $r=2m$ of the standard metric (see eq.
(\ref{eq:one}) of sect.\textbf{2}) is ``hidden'' in the
differentials of the new coordinates with respect to the standard
ones; $\beta$) \emph{an impairing} of the permanent gravitational
field of mass-point (a consequence of $\alpha$)); $\gamma$) a
time-dependent $\textrm{d}s^{2}$ (with the only exception of
metric \cite{1}, which is stationary, non-reversible) for a
\emph{static} problem.

\par The older \emph{maximally extended} metrics by Schwarzschild
\cite{5} and by Brillouin \cite{6} are generally ignored: a fact
that will be clarified by the future history of physics. We hope.

\vskip1.20cm \noindent \textbf{2.} -- The Eddington-Finkelstein
metric \cite{1} is obtained from the standard
(Hilbert-Droste-Weyl) metric

\begin{equation} \label{eq:one}
\textrm{d}s^{2} = \left(\frac{r}{r-\alpha}\right) \textrm{d}r^{2}
+ r^{2}\, \textrm{d}\omega^{2} - \left(\frac{r-\alpha}{r}\right)
\textrm{d}t^{2}
\end{equation}

-- where: $\alpha \equiv 2m$; $c=G=1$; $\textrm{d}\omega^{2}
\equiv \textrm{d}\vartheta^{2} + \sin^{2}\textrm{d}\varphi^{2}$ --
with the following transformation of time coordinate $t$:

\begin{equation} \label{eq:two}
t = t' \pm \alpha \ln |r-\alpha| \quad, \qquad \qquad \Rightarrow
\end{equation}

\begin{equation} \label{eq:twoprime}
\textrm{d}t = \textrm{d}t' \pm
\left(\frac{\alpha}{r-\alpha}\right) \, \textrm{d}r
 \quad; \tag{2\'{}}
\end{equation}

we see that both eqs. (\ref{eq:two}) and \ref{eq:twoprime})
contain the ``soft'' singularity at $r=\alpha$ of eq.
(\ref{eq:one}), that Finkelstein intended to remove from the
$\textrm{d}s^{2}$. We have:

\setcounter{equation}{2}
\begin{eqnarray} \label{eq:three}
\textrm{d}s^{2} & = & \left(\frac{r+\alpha}{r}\right)
\textrm{d}r^{2} + r^{2}\, \textrm{d}\omega^{2} -
\left(\frac{r-\alpha}{r}\right) \textrm{d}t'^{\,2} \mp
\left(\frac{2\alpha}{r}\right) \textrm{d}r \textrm{d}t' =
\nonumber\\
& & {} %%
= \textrm{d}r^{2} + r^{2}\, \textrm{d}\omega^{2} -
\textrm{d}t'^{\,2} \mp \frac{\alpha}{r} \, (\textrm{d}r +
\textrm{d}t')^{2} \quad;
\end{eqnarray}

the behaviour of the radial light-rays is particularly
interesting; from $0=\textrm{d}s^{2}= \textrm{d}\vartheta =
\textrm{d}\varphi$, we get two pairs of values for
$\frac{\textrm{d}r}{\textrm{d}t'}$:

\begin{equation} \label{eq:four}
\frac{r-\alpha}{r+\alpha}  \quad; \quad -1 \quad;
\end{equation}

\begin{equation} \label{eq:fourprime}
1 \quad; \quad -\frac{r-\alpha}{r+\alpha} \quad; \tag{4\'{}}
\end{equation}

light velocity \emph{depends} on the considered direction
(positive or negative) of coordinate $r$. Remark that eqs.
(\ref{eq:three}), (\ref{eq:four}), (\ref{eq:fourprime}) hold for
$r>0$. At $r=0$ we have a ``hard'' singularity (Kretschmann scalar
$=\infty$), the same of standard $\textrm{d}s^{2}$ of eq.
(\ref{eq:one}). Eqs. (\ref{eq:four}) and (\ref{eq:fourprime}) tell
us that light-rays do not ``feel'' the Hilbertian gravitational
repulsion \cite{7}; the non-reversibility of the
Eddington-Finkelstein field plays here an important role. And we
can also affirm that the radial geodesics of test-particle too are
not subjected to a Hilbertian repulsion. Transformation
(\ref{eq:two}) -- (\ref{eq:twoprime}) has sensibly impaired the
action of the gravitational field of mass $m$.

\par The ``hard'' singularity  at $r=0$ of eq. (\ref{eq:one}) can
be removed with a simple change of the radial coordinate $r$; for
instance:

\begin{equation} \label{eq:five}
r \rightarrow \left(r^{3}+\alpha^{3}\right)^{1/3} \quad,
\end{equation}

or

\begin{equation} \label{eq:fiveprime}
r \rightarrow r+\alpha \quad; \tag{5\'{}}
\end{equation}

with these substitutions, the form (\ref{eq:one}) yields,
respectively, the \emph{original} Schwarzschild form \cite{5} or
the Brillouin form \cite{6}. Quite analogously, (\ref{eq:five}),
or (\ref{eq:fiveprime}) remove the singularity at $r=0$ of eq.
(\ref{eq:three}). \emph{Physics remains unaltered!} Indeed, the
true physics of standard and Schwarzschild-Brillouin forms of
$\textrm{d}s^{2}$ concerns, respectively, the spatial regions
$r>\alpha$ and $r>0$; the distorted physics of eq.
(\ref{eq:three}), and of its transformed with substitution
(\ref{eq:five}) or (\ref{eq:fiveprime}), concerns, respectively,
the spatial regions $r>0$ and $r\geq 0$.

\vskip1.20cm \noindent \textbf{2bis.} -- Let us consider the pair
of expressions (\ref{eq:four}) and perform the integration of
$(\textrm{d}r/\textrm{d}t')=\left[(r-\alpha)/(r+\alpha)\right]$,
and of $(\textrm{d}r/\textrm{d}t')=-1$. We have

\begin{equation} \label{eq:sixone}
t' = r+2\alpha \ln|r-\alpha|+\textrm{const}    \quad;
\tag{6$_{1}$}
\end{equation}

\begin{equation} \label{eq:sixtwo}
t' = -r+\textrm{const}    \quad. \tag{6$_{2}$}
\end{equation}

According to (\ref{eq:sixone}), the light-rays which start from an
$r>\alpha$ move away from the origin $r=0$, while those which
start from an $r<\alpha$ go towards $r=0$, and reach it in a
finite time. According to (\ref{eq:sixtwo}),  all rays move
towards $r=0$, and reach it in a finite time. The pair of
expressions (\ref{eq:fourprime}) gives results that can be
obtained from eqs. (\ref{eq:sixone}) and (\ref{eq:sixtwo}) with
the substitution $t'\rightarrow -t'$. Remark that, in particular,
the equation
$(\textrm{d}r/\textrm{d}t')=-\left[(r-\alpha)/(r+\alpha)\right]$
tells us that the light-rays which start from an $r>\alpha$ will
reach $r=\alpha$ in an infinite time interval. Analogously, if in
eq. (\ref{eq:three}) we make, \emph{e.g.}, the substitution
(\ref{eq:fiveprime}), the integration of equation
$(\textrm{d}r/\textrm{d}t')=-\left[r/(r+2\alpha)\right]$ gives,
for a generic $\bar{r}$:

\begin{equation} \label{eq:sixprime}
\Delta t' = - \int_{\bar{r}}^{0}  \frac{r+2\alpha}{r} \,
\textrm{d}r= +\infty \quad. \tag{6\'{}}
\end{equation}

For the radial light-rays of \emph{standard} $\textrm{d}s^{2}$
(eq. (\ref{eq:one})), the integration of
$(\textrm{d}r/\textrm{d}t')=\pm \left[(r-\alpha)/r\right]$ gives:

\setcounter{equation}{6}
\begin{equation} \label{eq:seven}
t = \pm \Bigg[r+\alpha
\ln\left|\frac{r-\alpha}{\alpha}\right|\Bigg] +\textrm{const}
\quad, \quad (r>\alpha) \quad;
\end{equation}

the surface $r=\alpha$ is reached in an \emph{infinite} time
interval (an instance of Hilbertian repulsion).

\vskip1.20cm \noindent \textbf{3.} -- The Lema\"itre
time-dependent metric \cite{2} is obtained from eq. (\ref{eq:one})
by means of the following coordinate transformations:

\begin{equation} \label{eq:eight}
r = \left[\frac{3}{2} \,  \alpha^{1/2} \, (\tau-\chi)\right]^{2/3}
\quad;
\end{equation}

\begin{equation} \label{eq:eightprime}
t = \tau +2\,(\alpha \, r)^{1/2} + \alpha \, \ln \left|
\frac{r^{1/2}-\alpha^{1/2}}{r^{1/2}+\alpha^{1/2}}\right| \quad;
\tag{8\'{}}
\end{equation}

from which:

\begin{equation} \label{eq:eightsecond}
\textrm{d}t = \textrm{d}\tau + \frac{r}{r-\alpha} \, \alpha^{1/2}
\,  r^{-1/2} \textrm{d}r \quad. \tag{8\'{}\'{}}
\end{equation}

We have:

\setcounter{equation}{8}
\begin{equation} \label{eq:nine}
\textrm{d}s^{2} = \frac{\alpha}{r} \, \textrm{d}\chi^{2} + r^{2}
\, \textrm{d}\omega^{2} - \textrm{d}\tau^{2}\quad.
\end{equation}

The speed of the radial $(\textrm{d}\omega =0)$ light-rays is:

\begin{equation} \label{eq:ten}
\frac{\textrm{d}\chi}{\textrm{d}\tau} = \pm \left(
\frac{r}{\alpha}\right)^{1/2}  \quad;
\end{equation}

thus: $\textrm{d}\chi / \textrm{d}\tau =0$ for $r=0$;
$\textrm{d}\chi / \textrm{d}\tau =\pm 1$ for $r=\alpha$;
$\textrm{d}\chi / \textrm{d}\tau =\pm \infty$ for $r=\infty$. As
it is clear, there is a Hilbertian repulsion along the whole
trajectory (as it happens for eq. (\ref{eq:one}), cfr. eq.
(\ref{eq:seven})). And the test-particles moving along radial
geodesics will ``feel'' the Hilbertian repulsion in some portions
of their trajectories (as it happens for eq. (\ref{eq:one})).

\par The ``hard'' singularity at $r=0$ of eq. (\ref{eq:nine}) can
be removed by one (\emph{ad libitum}) of the substitutions
(\ref{eq:five}), (\ref{eq:fiveprime}), exactly as in the case of
Eddington-Finkelstein form of $\textrm{d}s^{2}$.

\par Robertson metric \cite{3} can be obtained from Lema\"itre
metric with the transformation

\begin{equation} \label{eq:eleven}
\chi = -\frac{2}{3} \, \chi'^{3/2} \alpha^{-1/2} \quad;
\end{equation}

clearly, this metric has the same general properties of
Lema\"itre's one.

\vskip1.20cm \noindent \textbf{4.} -- The interval
$\textrm{d}s^{2}$ of Kruskal-Szekeres metric is \cite{4}:

\begin{equation} \label{eq:twelve}
\textrm{d}s^{2} = \frac{4\, \alpha^{3}}{r} \,
\exp\left(-\frac{r}{\alpha}\right)\,
(\textrm{d}u^{2}-\textrm{d}v^{2}) + r^{2}\textrm{d}\omega^{2}
\quad;
\end{equation}

where $r$ is a function of the space-like coordinate $u$,
$(-\infty < u < +\infty)$, and of the time-like coordinate $v$,
$(-\infty < v < +\infty)$; more precisely:

\begin{equation} \label{eq:thirteen}
\left(\frac{r}{\alpha}-1\right) \, \exp
\left(\frac{r}{\alpha}\right) = u^{2}-v^{2} \quad; \quad
\textrm{and}
\end{equation}

\begin{equation} \label{eq:thirteenprime}
t = 2\alpha \, \textrm{arctanh} \left(\frac{v}{u}\right) \quad.
\tag{13\'{}}
\end{equation}

This metric is invariant under the substitutions $u\rightarrow -u$
and $v\rightarrow -v$. Each point of metric (\ref{eq:one}) has a
twofold representation in metric (\ref{eq:twelve}): an odd-looking
\emph{embarras de richesse}.

\par Seemingly, the singularity at $r=\alpha$ of eq.
(\ref{eq:one}) does not appear in eq. (\ref{eq:twelve}). Now, the
differentials $\textrm{d}u$, $\textrm{d}v$ of the functions
$u(r,t)$, $v(r,t)$ are singular at $r=\alpha$ ! (See sect.
\textbf{A3} of the Appendix).

\par Metric (\ref{eq:twelve}) suffers from various defects, for
instance: \emph{i}) it is $v$-dependent, \emph{i.e.} dependent on
a time-like coordinate, \emph{ii}) the radial
($\textrm{d}\omega=0$) light-rays do \emph{not} ``feel'' the
gravity: $\textrm{d}s^{2} =0$  gives $\textrm{d}u= \pm \,
\textrm{d}v$, the light-cones are ``open'' as in \emph{special}
relativity: an apparent worth, a physical fault, a revenge of the
``soft'' singularity at $r=\alpha$ of eq. (\ref{eq:one}), which
has been swept away into a commonly unobserved corner (the
differentials $\textrm{d}u$, $\textrm{d}v$).

\vskip1.20cm \noindent \textbf{4bis.} -- The ``hard'' singularity
at $r=0$ of metric (\ref{eq:twelve}) can be removed with a
suitable substitution of the radial coordinate, \emph{e.g.} with
(\ref{eq:five}) or (\ref{eq:fiveprime}). The \emph{new} $r=0$
represents the \emph{previous} $r=\alpha$; the interior region
$r<\alpha$ loses any meaning and dies away from existence: a
trivial consequence of the fact that the choice of the radial
coordinate is quite free, and allows a \emph{shifting} of the
standard $r$, which eliminates the ``hard'' singularity at $r=0$,
in spite of the \emph{infinite} value of its Kretschmann scalar.
(Physics does not always coincide with geometry).

\par If, for instance, we perform the shifting $r\rightarrow
r+\alpha$, write for clarity's sake $r=\varrho+\alpha$, and call
$U$, $V$ the new space-like and time-like coordinates, eq.
(\ref{eq:twelve}) becomes:

\begin{equation} \label{eq:twelveprime}
\textrm{d}s^{2} = -\frac{4\,\alpha^{3}}{\varrho+\alpha} \,
\exp\left(-\frac{\varrho+\alpha}{\alpha}\right)\,
(\textrm{d}V^{2}-\textrm{d}U^{2}) +
(\varrho+\alpha)^{2}\textrm{d}\omega^{2} \quad; \quad (0\leq
\varrho <\infty) \quad . \tag{12\'{}}
\end{equation}

We have:

\begin{displaymath} \label{eq:fourteen}
\left\{ \begin{array}{l} U(\varrho,t) =  \left(\displaystyle
\frac{\varrho}{\alpha}\right)^{1/2} \exp
\left(\displaystyle\frac{\varrho+\alpha}{2\alpha}\right) \, \cosh
\left(\displaystyle
\frac{t}{2\alpha}\right) \quad, \\ \\
V(\varrho,t) =  \left(\displaystyle
\frac{\varrho}{\alpha}\right)^{1/2} \exp
\left(\displaystyle\frac{\varrho+\alpha}{2\alpha}\right) \, \sinh
\left(\displaystyle \frac{t}{2\alpha}\right) \quad; \tag{14}
\end{array} \right.
\end{displaymath}

from which:

\begin{equation} \label{eq:fourteenprime}
\left( \frac{\varrho}{\alpha}\right) \exp
\left(\frac{\varrho+\alpha}{\alpha}\right)= U^{2}-V^{2} \quad.
\tag{14\'{}}
\end{equation}

\begin{equation} \label{eq:fourteensecond}
t = 2\,\alpha \, \textrm{arctanh} \left(\frac{V}{U}\right) \quad.
\tag{14\'{}\'{}}
\end{equation}

Apart from the substitutions $U\rightarrow -U$ and $V\rightarrow
-V$, we have here a \emph{unique} form (\ref{eq:fourteen}) for the
functions $U(\varrho,t)$, $V(\varrho,t)$. On the contrary, in the
Kruskal-Szekeres metric (\ref{eq:twelve}) there are four different
pairs of coordinates: two for $r>\alpha$, and two for $r<\alpha$:
a real patchwork. (See sect.\textbf{A3} of the Appendix).

\vskip1.20cm \noindent \textbf{4ter.} -- A constant radial
coordinate, $r=\textrm{const}$, is represented in a Cartesian
plane $(u,v)$ -- or $(U,V)$ -- by an equilateral hyperbola
$u^{2}-v^{2}=\textrm{const}$ -- or $U^{2}-V^{2}=\textrm{const}$.
If we choose as new coordinates, say $(u',v')$ -- or $(U',V')$,
the asymptotes of these hyperbolae, their equations become
$u'v'=\textrm{const}$ -- or $U'V'=\textrm{const}$, but we have
lost the difference between space-like and time-like coordinates:
a not negligible disadvantage, from a physical standpoint. The
null lines of radial ($\textrm{d}\omega=0$) light-rays are
represented by equations $u'=\textrm{const}$, $v'=\textrm{const}$
-- or $U'=\textrm{const}$, $V'=\textrm{const}$.

\par Some authors take the equations $u'=\textrm{const}$,
$v'=\textrm{const}$ as a starting point for a direct derivation of
Kruskal-Szekeres metric, avoiding any reference to eq.
(\ref{eq:one}). Thus, they \emph{postulate} that the radial
light-rays are \emph{not} subjected to the gravitational field of
mass $m$. This is an \emph{ad hoc} assumption: \emph{ad hoc},
because its aim is the exclusion from the metric of the ``soft''
singularity at $r=\alpha$ (finite value of Kretschmann scalar).
Now, this singularity, which could be qualified as non-physical
because the mass-point is in $r=0$, is not a mere spurious
hindrance: as it was \emph{first} demonstrated by Schwarzschild
\cite{5} in the \emph{original} construction of the homonymous
manifold, it plays actually a fundamental role. Indeed
Schwarzschild $\textrm{d}s^{2}$ is:

\setcounter{equation}{14}
\begin{equation} \label{eq:fifteen}
\textrm{d}s^{2} =
\left(\frac{\textrm{R}}{\textrm{R}-\alpha}\right) \textrm{dR}^{2}
+
\textrm{R}^{2}\textrm{d}\omega^{2}-\left(\frac{\textrm{R}-\alpha}{\textrm{R}}\right)\textrm{d}t^{2}
\quad,
\end{equation}

where: $\textrm{R}\equiv (r^{3}+\alpha^{3})^{1/3}$; $0<r<\infty$.
The point-mass $m$ is situated in $r=0$, see the correspondence
with Newton theory; the singularity at $r=0$ of eq.
(\ref{eq:fifteen}) coincides with the singularity at $r=\alpha$ of
eq. (\ref{eq:one}).

\par When we look for a solution of Einstein equations $R_{jk}=0$,
$(j, k=1, 2, 3, 4)$, -- a solution \emph{with singularities}, we
mean --, we must be very careful about the choice of the reference
system. Indeed, a system which appears simple and reasonable from
a \emph{geometrical} standpoint, can originate some misleading
properties, as for instance a weakening of the permanent
gravitational fields.

\par Kruskal-Szekeres metric \cite{4} ``does not make physical
sense'', as Bonnor wrote in the article quoted in \cite{4}. and a
similar negative judgement was expressed by this author on the
Novikov metric (1963) of Schwarzschild manifold, which ``throws
some light on the Kruskal diagram $[(u,v)]$, without removing all
its obscurities.''

\vskip1.20cm \noindent \textbf{Conclusion} -- We have evidenced
the shortcomings of the metrics by Eddington-Finkelstein,
Lema\"itre and Robertson, Kruskal-Szekeres. Schwarzschild's
\emph{original} $\textrm{d}s^{2}$ \cite{5} and Brillouin's
$\textrm{d}s^{2}$ \cite{6} give maximally extended metrics which
describe perfectly the physical reality, and make clear that
standard $\textrm{d}s^{2}$ (eq. (\ref{eq:one})) holds \emph{only
for} $r>\alpha$. A fact confirmed by Hilbertian gravitational
repulsion \cite{7}.

% \newpage
\vskip2.00cm
\begin{center}
\noindent \small \emph{\textbf{APPENDIX}} \normalsize
\\ ``\emph{Bildr\"aume'' (Weyl) and representative spaces (Synge)}
\end{center}

\vskip0.40cm \noindent \textbf{A1.} -- The notion of
\emph{Bildraum} (picture space) has been introduced in GR by Weyl
(see, \emph{e.g.}, \cite{8}). Synge spoke of a ``representative
space'', theorized its use and applied it in his study on ``The
gravitational field of a particle'' \cite{2}. In a particular and
important problem, Fock employed a ``conformal space'' \cite{9}.
Eddington utilized the concept in a subtle and indirect way when
he wrote, at the beginning of his treatment of Schwarzschild
manifold \cite{10}: ``In a flat space-time the interval, referred
to spherical polar coordinates and time, is $\textrm{d}s^{2}=-
\textrm{d}r^{2}-r^{2}\textrm{d}\vartheta^{2}-r^{2}\sin^{2}\vartheta\textrm{d}\varphi^{2}+\textrm{d}t^{2}$.
-- If we consider what modifications of this can be made without
destroying the spherical symmetry in space, the symmetry as
regards past and future time, or the static condition, the most
general possible form appears to be $\textrm{d}s^{2}=-
U(r)\textrm{d}r^{2}-V(r)
(r^{2}\textrm{d}\vartheta^{2}+r^{2}\sin^{2}\vartheta\textrm{d}\varphi^{2})+W(r)\textrm{d}t^{2}$.''
Boyer and Lindquist, in their study of Kerr's metric \cite{11},
introduced a ``Euclidean 3-space with Cartesian coordinates''.

\par The great majority of the authors utilize implicitly representative spaces,
and often without a clear distinction between features of the
considered space-time of GR and features of its picture space.

\vskip1.20cm \noindent \textbf{A2.} -- In sect.\textbf{2} of paper
\cite{2} Synge wrote: ``Once we have decided on the idealized
experiments which we shall use, we have thereby set up \emph{a
system of coordinates} $x^{r}$ \emph{in a space-time}. $[$...$]$.
The next step is to make a \emph{geometrical representation} of
space-time. $[$...$]$. We think then of a space $V_{4}$ of four
dimensions -- a \emph{representative space}. $[$...$]$. The
representative space is a \emph{map} of space-time, and like every
map it is a mixture of the intrinsic properties of the thing
mapped and certain conventionalities introduced for our human
convenience in understanding and interpreting. $[$...$]$.
Modifications for convenience may be introduce later, but let us
start with the idea that our representative space $V_{4}$ is a
Euclidean space of four dimensions.'' In sect.\textbf{3} of
\cite{2} we read: ``Let $(u,v)$ be two variables ranging from
$-\infty$ to $+\infty$. They will be taken, for purposes of
representation, as rectangular Cartesians in a Euclidean plane
$U_{2}.$'' And in sect.\textbf{4} of \cite{2}: ``The rest of that
section $[$\emph{i.e.}, of sect.\textbf{3}$]$ was devoted to the
definition of certain functions of $(u^{2}-v^{2})$. Among these
functions was $r$  $[$...$]$. The plane $U_{2}$ forms half of our
representation. The other half is provided through a family of
concentric spheres on which the variables $\vartheta$ and
$\varphi$ are respectively colatitude and azimuth referred to a
common pole $\vartheta=0$ and to a common base plane $\varphi=0$.
$[$...$]$. -- To the assigned pair of values $(u,v)$ there
corresponds a point $P$ in the plane $U_{2}$ and also $[$...$]$ a
value of $r$ in the range $0\leq r<+\infty$. Hence there
corresponds a sphere $S_{2}$ of radius $r$ in the above mentioned
concentric family. Assigned values of $(\vartheta, \varphi)$ fix a
point $Q$ on $S_{2}$. $[$...$]$. \emph{We shall define our
representative space} $V_{4}$ \emph{by saying that a point of}
$V_{4}$ \emph{is a point-pair} $(P, Q)$. $[$...$]$. So far nothing
of space-time. $[$...$]$. Hypothesis A: \emph{All events in
space-time containing a single gravitating particle may be put in
one to one correspondence with the points of the representative
space} $V_{4}$ \emph{described above}.

\par As regards the line-element of space-time, let us set down for consideration the form
$\Phi=
\textrm{d}u^{2}-\textrm{d}v^{2}+(v\,\textrm{d}u-u\,\textrm{d}v)^{2}
F
+r^{2}(\textrm{d}\vartheta^{2}+\sin^{2}\vartheta\textrm{d}\varphi^{2})$,
$F$ and $r$ being functions of $(u^{2}-v^{2}$ as defined in
Section 3, these functions involving a positive constant
$a\,[\equiv2\,m]$.''

\par The above $\Phi$, \emph{i.e.} $\textrm{d}s^{2}$, is the
\emph{clou} of a complex investigation, which inspired Kruskal
\cite{4} and Szekeres \cite{4}, who succeeded in giving a
simplified and more manageable version of Synge's results.

\vskip1.20cm \noindent \textbf{A3.} -- Back to Kruskal-Szekeres
metric. The representative space is identical to Synge's one: a
Euclidean plane $U_{2}$, referred to Cartesian orthogonal axes
$(u,v)$, and a set $S_{2}$ of concentric spheres on which a
colatitude $\vartheta$ and an azimuth $\varphi$ are defined.
Metric (\ref{eq:twelve}) is referred to four different pairs of
coordinates; accordingly, representative plane $U_{2}$ is divided
into four regions I, II, III, IV.

\begin{displaymath} \label{eq:A1}
\left\{ \begin{array}{l} u_{I} = \left(\displaystyle
\frac{r}{\alpha}-1\right)^{1/2} \exp
\left(\displaystyle\frac{r}{2\alpha}\right) \, \cosh
\left(\displaystyle
\frac{t}{2\alpha}\right) \quad; \\ \\
v_{I} = \left(\displaystyle \frac{r}{\alpha}-1\right)^{1/2} \exp
\left(\displaystyle\frac{r}{2\alpha}\right) \, \sinh
\left(\displaystyle \frac{t}{2\alpha}\right) \quad, \tag{A$_{1}$}
\end{array} \right.
\end{displaymath}

for $r>\alpha$, and

\begin{displaymath} \label{eq:A2}
\left\{ \begin{array}{l} u_{II} = \left(\displaystyle
1-\frac{r}{\alpha}\right)^{1/2} \exp
\left(\displaystyle\frac{r}{2\alpha}\right) \, \sinh
\left(\displaystyle \frac{t}{2\alpha}\right) \quad; \\ \\ v_{II} =
\left(\displaystyle1- \frac{r}{\alpha}\right)^{1/2} \exp
\left(\displaystyle\frac{r}{2\alpha}\right) \, \cosh
\left(\displaystyle \frac{t}{2\alpha}\right) \quad, \tag{A$_{2}$}
\end{array} \right.
\end{displaymath}

for $r<\alpha$. From (\ref{eq:A1})--(\ref{eq:A2}) we have, in
particular:

\begin{displaymath} \label{eq:A3}
\left( \frac{r}{\alpha}-1\right) \exp
\left(\frac{r}{\alpha}\right) = \left\{ \begin{array}{l}
u_{I}^{2}-v_{I}^{2} \quad; \\ \\ u_{II}^{2}-v_{II}^{2}  \quad.
\tag{A$_{3}$}
\end{array} \right.
\end{displaymath}

The Cartesian plane $(u,v)$ is divided into four quadrants I, II,
III, IV by the null lines $u=v$, $u=-v$, that are null lines of
light-rays. Right-hand quadrant I and upper quadrant II cover the
entire Schwarzschild space-time; left-hand quadrant III is a
pendant of I: $u_{III}=-u_{I}$, $v_{III}=-v_{I}$; lower quadrant
IV is a pendant of II: $u_{IV}=-u_{II}$, $v_{IV}=-v_{II}$. Formula
(\ref{eq:A3}) holds also for quadrants III and IV. (Formula
(\ref{eq:thirteenprime}) of sect.\textbf{4} holds too for all
quadrants).

\par The above patchwork is not only redundant, because quadrants
I and II are sufficient to describe both the exterior and the
interior regions of $r=\alpha$, but has also this \emph{surprising
property}: if we substitute in metric (\ref{eq:twelve}) any
whatever of the four coordinate pairs $(u_{I}, v_{I})$, $(u_{II},
v_{II})$, $(u_{III}, v_{III})$, $(u_{IV}, v_{IV})$,, we obtain
always the standard $\textrm{d}s^{2}$ of eq. (\ref{eq:one}),
\emph{without} any distinction between the exterior and the
interior region of surface $r=\alpha$.

\par Kruskal-Szekeres metric is a good example of the heuristic
and interpretative value of a convenient \emph{Bildraum}. However,
Synge's representative space and Kruskal-Szekeres metric do not
give a \emph{faithful} description of physical reality, owing to
the defects that we have pointed out in sect.\textbf{4}.

 \vskip1.20cm \noindent \textbf{A4.} -- In sect.\textbf{A1} we
 have written that, in his construction of Schwarzschild manifold,
 Eddington utilized the concept of \emph{Bildraum} in a subtle and
 indirect way. Indeed, for the spherical symmetry in space-time of
 GR he drew inspiration from the Minkowskian $\textrm{d}s^{2}$
 expressed with spherical polar coordinate, and wrote
 $\textrm{d}s^{2}=-
U(r)\textrm{d}r^{2}-V(r)
(r^{2}\textrm{d}\vartheta^{2}+r^{2}\sin^{2}\vartheta\textrm{d}\varphi^{2})+W(r)\textrm{d}t^{2}$.
We emphasize that \emph{in GR} the notion of spherical symmetry is
not a well defined and understood concept (Synge \cite{2}). Thus,
Eddington (as -- more or less implicitly -- all the Fathers of
Relativity) took advantage of the fact that, on the contrary,
spherical symmetry can be perfectly mastered \emph{in SR}. Then,
he wrote: $r^{2}\, V(r) \rightarrow r^{2}$: and we can say that
the free choice of the radial coordinate allowed him to exploit
Synge's family $S_{2}$ of concentric spheres. Finally, at p.94 of
\cite{10}, our Author pointed out that the general solution in
spherical polar coordinates of Schwarzschild problem can be
obtained by substituting the $r$ of standard form of solution (eq.
(\ref{eq:one})) with \emph{any} regular function $f(r)$. A result
that can be recovered by solving equations $R_{jk}=0$ for
$g_{rr}=-U(r)$; $g_{\vartheta\vartheta}=-V(r)\,r^{2}$;
$g_{\varphi\varphi}=-V(r)\,r^{2}\sin^{2}\vartheta$; $g_{tt}=W(r)$
-- see, \emph{e.g.}, the Appendix of Abrams \cite{12}.

\par The instance of Schwarzschild manifold is emblematic: as a
matter of fact, in all problems of GR the pseudo-Riemannian
manifold is not known \emph{a priori}, it is obtained by solving
the concerned Einsteinian equations. Consequently, the starting
point of the investigation is always the (implicit or explicit)
consideration of a \emph{Bildraum}, that we choose taking heed of
the general properties (\emph{e.g.}, spherical symmetry) of our
problem.


\vskip1.80cm \small
\begin{thebibliography}{99}

\bibitem{1}
A. S. Eddington, \emph{Nature}, \textbf{113} (1924) 192 -- in
eq.(2) there is a trivial misprint: $m$ in lieu of $2m$; D.
Finkelstein, \emph{Phys. Rev.}, \textbf{110} (1958) 965. See also
L. Landau et E. Lifchitz, \emph{Th\'eorie du Champ}, Deuxi\`eme
\'edition revue (\'Editions MIR, Moscou) 1966, sect.\textbf{97}.


\bibitem{2}
G. Lema\"itre, \emph{Ann. Soc. Sci. Bruxelles}, \textbf{53A}
(1933) 51. See also: J.L. Synge, \emph{Proc. Roy. Irish Acad.},
\textbf{53A} (1950) 83; C. M\o ller, \emph{The Theory of
Relativity}, Second Edition (Clarendon Press, Oxford) 1972, p.442;
P.A.M. Dirac, \emph{General Theory of Relativity} (J. Wiley and
Sons, New York, \emph{etc.}) 1975, sect.\textbf{19}.

\bibitem{3}
H. P. Robertson, see p.84 of Synge \cite{2}.

\bibitem{4}
M. Kruskal, \emph{Phys. Rev.}, \textbf{119} (1960) 1743; Gy.
Szekeres, \emph{Publ. Mat. Debrecen}, \textbf{7} (1960) 285; the
long and laborious work by Synge quoted in \cite{2} is an ancestor
of the papers by Kruskal and Szekeres. See further the review
article by W. B. Bonnor in \emph{Gen. Rel. Grav.}, \textbf{24}
(1992) 551.

\bibitem{5}
K. Schwarzschild, \emph{Berl. Ber.}, (1916) 189; for an English
version, see \emph{arXiv:physics/9905030}, May 12th, 1999 -- and
\emph{Gen. Rel. Grav.}, \textbf{35} (2003) 951. See further S.
Antoci and D.-E. Liebscher, \emph{Astr. Nachr.}, \textbf{322}
(2001) 137 -- and references therein.

\bibitem{6}
M. Brillouin, \emph{Journ. Phys. Rad.}, \textbf{23} (1923) 43; for
an English version, see \emph{arXiv:physics/0002009}, February
3rd, 2000.

\bibitem{7}
D. Hilbert, \emph{Mathem. Annalen}, \textbf{92} (1924) 1; also in
\emph{Gesammelte Abhandlungen}, Dritter Band (J. Springer, Berlin)
1935, p.258. See further A. Loinger and T. Marsico,
\emph{arXiv:0904.1578 v1} $[$physics.gen-ph$]$ 9 Apr 2009 -- and
references therein.

\bibitem{8}
H. Weyl, \emph{Raum-Zeit-Materie}, Siebente Auflage
(Springer-Verlag, Berlin, \emph{etc.}) 1988, sects. \textbf{33},
\textbf{35}, \textbf{37}.

\bibitem{9}
V. Fock, \emph{The Theory of Space, Time and Gravitation}, 2nd
Revised Edition (Pergamon Press, Oxford, \emph{etc}.) 1964,
pp.203-204.

\bibitem{10}
A. S. Eddington, \emph{The Mathematical Theory of Relativity},
Second Edition (Cambridge University Press, Cambridge) 1960, p.83.

\bibitem{11}
R.H. Boyer and R.W Lindquist, \emph{J. Math. Phys.}, \textbf{8}
(1967) 265.

\bibitem{12}
L.S. Abrams, \emph{Phys. Rev.}, \textbf{20} (1979) 2474; also in
\emph{arXiv:gr-qc/0201044 v1} (14 Jan 2002). In this interesting
paper, the Author empasizes the great physical value of
Schwarzschild's original solution \cite{5} -- a value that is
generally overlooked by the current literature.

\end{thebibliography}
\end{document}